\documentclass[prl,showpacs,aps,twocolumn,floats]{revtex4}
\usepackage{bbm,epsfig,citesort}

\newcommand{\p}{\partial}

\begin{document}
\title{Rotor Spectra, Berry Phases, and Monopole Fields: 
from Antiferromagnets to QCD}
\author{S.\ Chandrasekharan$^a$, F.-J.\ Jiang$^b$, M.\ Pepe$^c$, and 
U.-J.\ Wiese$^b$}
\affiliation {$^a$ Department of Physics, Box 90305, Duke University, Durham, 
North Carolina 27708 \\
$^b$ Institute for Theoretical Physics, Bern University,
Sidlerstrasse 5, 3012 Bern, Switzerland \\
$^c$ Istituto Nazionale di Fisica Nucleare and
Dipartimento di Fisica, \\
Universit\`a di Milano-Bicocca, 3 Piazza della Scienza, 20126 Milano, Italy}


\begin{abstract}
The order parameter of a finite system with a spontaneously broken continuous 
global symmetry acts as a quantum mechanical rotor. Both antiferromagnets with 
a spontaneously broken $SU(2)_s$ spin symmetry and massless QCD with a broken
$SU(2)_L \times SU(2)_R$ chiral symmetry have rotor spectra when considered in 
a finite volume. When an electron or hole is doped into an antiferromagnet or 
when a nucleon is propagating through the QCD vacuum, a Berry phase arises from
a monopole field and the angular momentum of the rotor is quantized in 
half-integer units.
\end{abstract} 

\maketitle

Berry phases and monopole fields are familiar from adiabatic processes in
quantum mechanical systems \cite{Ber84,Sim83}. For example, the slow rotation 
of the nuclei in a diatomic molecule is influenced by a geometric vector 
potential generated by the fast motion of the electrons \cite{Her63,Mea80}. The
Abelian and non-Abelian monopole content of these vector potentials was worked 
out elegantly by Moody, Shapere, and Wilczek \cite{Moo86}. In this paper we 
discuss Berry phases and monopole fields for rotors arising in condensed matter
and particle physics systems with a spontaneously broken continuous global 
symmetry.

The undoped precursors of layered cuprate high-temperature superconductors are 
antiferromagnets with a spontaneously broken $SU(2)_{s}$ spin symmetry. When 
one considers an antiferromagnet of finite volume $V$ at very low temperatures,
the dynamics are dominated by the spatially independent zero-mode of the 
staggered magnetization order parameter
\begin{equation}
\vec{e}(t) = (\sin\theta(t) \cos\phi(t),\sin\theta(t) \sin\phi(t),
\cos\theta(t)),
\end{equation}
which represents a slow quantum mechanical rotor governed by the Lagrangian 
\cite{Has93}
\begin{equation}
{\cal L} = \frac{\Theta}{2} \p_t \vec e \cdot \p_t \vec e =
\frac{\Theta}{2} \left[(\p_t \theta)^2 + \sin^2\theta (\p_t\varphi)^2\right].
\end{equation}
Integrating out the fast non-zero modes of the staggered magnetization at one 
loop, and assuming a 2-dimensional quadratic periodic volume, the moment
of inertia was determined by Hasenfratz and Niedermayer \cite{Has93} as
\begin{equation}
\Theta = \frac{\rho_s V}{c^2}
\left[1 + \frac{3.900265}{4 \pi} \left(\frac{c}{\rho_s L}\right) +
{\cal O}\left(\frac{1}{L^2}\right) \right],
\end{equation}
where $\rho_s$ is the spin stiffness and $c$ is the spinwave velocity. The 
momenta conjugate to $\theta$ and $\varphi$ are
\begin{equation}
p_\theta = \frac{\delta {\cal L}}{\delta \p_t\theta} = \Theta \ \p_t\theta, \
p_\varphi = \frac{\delta {\cal L}}{\delta \p_t\varphi} = 
\Theta \ \sin^2\theta \ \p_t\varphi,
\end{equation}
and the resulting Hamiltonian
\begin{equation}
H = - \frac{1}{2 \Theta} 
\left(\frac{1}{\sin\theta} \p_\theta [\ \sin\theta \p_\theta] + 
\frac{1}{\sin^2\theta} \p_\varphi^2\right) = \frac{\vec L^2}{2 \Theta} 
\end{equation}
is just the Laplacian on the sphere $S^2$. Correspondingly, the energy spectrum
is that of a quantum mechanical rotor with angular momentum 
$l \in \{0,1,2,...\}$, i.e.
\begin{equation}
\label{Energy}
E_l = \frac{l(l+1)}{2 \Theta},
\end{equation}
with each state being $(2l+1)$-fold degenerate. The rotor features have been
verified in numerical simulations of the antiferromagnetic quantum Heisenberg 
model \cite{Wiese92,Bea96} (or equivalently of the $t$-$J$ model at half-filling). It 
should be noted that a quantum ferromagnet does not behave as a rotor because 
its order parameter --- the uniform magnetization --- is a conserved quantity.

When a single hole or electron is doped into the antiferromagnet, the spin of 
the system changes by $1/2$ and thus the angular momentum of the resulting 
rotor must then be quantized in half-integer units. As we will see, in the 
language of low-energy effective theories, this half-integer quantization is a
result of Berry phases and monopole fields. Systematic
low-energy effective theories for charge carriers in an antiferromagnet were 
recently constructed in \cite{Kae05,Bru06}. The leading terms in the low-energy
Lagrangian of holes or electrons with a small momentum $\vec p$ are given by
\begin{equation}
{\cal L} = \frac{\Theta}{2} \p_t \vec e \cdot \p_t \vec e \nonumber \\
+ \Psi^\dagger \left[E(\vec p) - i \p_t + v_t^3 \sigma_3 + 
\lambda V_t \right] \Psi.
\end{equation}
Here 
$\Psi(t) = \left(\begin{array}{c} \psi_+(t) \\ \psi_-(t) \end{array}\right)$ is
a two-component Grassmann valued field describing fermions with spin parallel 
($+$) or anti-parallel ($-$) to the local staggered magnetization. It should be
noted that we have suppressed an additional flavor index of the hole fields in
a doped cuprate antiferromagnet \cite{Bru06}, which distinguishes between holes
from  different pockets in the Brillouin zone. The fermion 
energy $E(\vec p)$ as well as $\lambda$ can be determined by integrating out 
the non-zero momentum modes of the staggered magnetization, e.g.\ at one loop. 
For hole- or electron-doped cuprates as well as for the $t$-$J$ model it was
predicted that $\lambda = 0$ \cite{Bru06}, while for other antiferromagnets in 
general $\lambda \neq 0$ \cite{Kae05}. The Abelian vector potential $v_t^3(t)$ 
is the diagonal component of the composite vector field
\begin{equation}
v_t = u \p_t u^\dagger = i v_t^a \sigma_a = i v_t^3 \sigma_3 + i V_t.
\end{equation}
Here $\sigma_a$ are the Pauli matrices and
\begin{equation}
u = \left(\begin{array}{cc} \cos\frac{\theta}{2} &
\sin\frac{\theta}{2} \exp(- i \varphi) \\ 
- \sin\frac{\theta}{2} \exp(i \varphi) &
\cos\frac{\theta}{2} \end{array}\right)
\end{equation}
represents a transformation which rotates $\vec e(t)$ into the $3$-direction.
One then obtains
\begin{eqnarray}
v_t^3&=&\sin^2\frac{\theta}{2} \p_t \varphi, \nonumber \\
V_t&=&\frac{1}{2} \sin\theta 
(\cos\varphi \ \sigma_1 + \sin\varphi \ \sigma_2) \ \p_t \varphi
\nonumber \\
&+&\frac{1}{2}
(\ \sin\varphi \ \sigma_1 - \cos\varphi \ \sigma_2) \ \p_t \theta.
\end{eqnarray} 
These velocity-dependent terms give rise to a modification of the canonically
conjugate momenta such that
\begin{equation}
\Theta \ \p_t \theta = p_\theta + i A_\theta, \
\Theta \ \p_t \varphi = \frac{1}{\sin^2\theta}(p_\varphi + i A_\varphi),
\end{equation}
with the non-Abelian vector potential
\begin{eqnarray}
A_\theta&=&i\frac{\lambda}{2} (\sin\varphi \ \sigma_1 - 
\cos\varphi \ \sigma_2), \nonumber \\
A_\varphi&=&i\sin^2\frac{\theta}{2} \sigma_3 +
i \frac{\lambda}{2} \sin\theta (\cos\varphi \ \sigma_1 + 
\sin\varphi \ \sigma_2), \nonumber \\
\end{eqnarray}
and the corresponding field strength
\begin{equation}
\label{F}
F_{\theta\varphi} = \p_\theta A_\varphi - \p_\varphi A_\theta + 
[A_\theta,A_\varphi] = i \frac{1 - \lambda^2}{2} \sin\theta \ \sigma_3.
\end{equation} 
Remarkably, the resulting geometric Berry gauge field is exactly the same as
for a diatomic molecule \cite{Moo86}. For cuprates ($\lambda = 0$) the vector
potential is Abelian and describes a monopole with quantized magnetic flux. The
path $\vec e(t)$ in periodic Euclidean time defines a closed loop ${\cal C}$ on
$S^2$. The Boltzmann factor in the path integral contains a Wilson loop along 
${\cal C}$ which manifests itself as a Berry phase. Using Stokes' theorem, the 
Berry phase is given by the magnetic flux enclosed in ${\cal C}$. Since the 
enclosed flux is well-defined only up to the area $4 \pi$ of $S^2$, the 
magnetic charge is $\pm \frac{1}{2}$ as a consequence of the Dirac quantization
condition. For a general antiferromagnet ($\lambda \neq 0$) the vector 
potential becomes non-Abelian and the flux is no longer quantized.

The resulting Hamilton operator takes the form
\begin{eqnarray}
H(\lambda)&=&- \frac{1}{2 \Theta} \left\{\frac{1}{\sin\theta}
(\p_\theta - A_\theta)[\ \sin\theta (\p_\theta - A_\theta)] \right.
\nonumber \\
&+&\left.\frac{1}{\sin^2 \theta}(\p_\varphi - A_\varphi)^2\right\} +
E(\vec p).
\end{eqnarray}
The solution for the energy spectrum can be obtained along the lines of 
\cite{Moo86}. First, one can show that the Hamiltonian $H(0)$ (with 
$\lambda = 0$) commutes with the angular momentum operators
\begin{eqnarray}
\label{J}
&&J_\pm = \exp(\pm i \varphi) \left(\pm \ \p_\theta + i \cot\theta \ \p_\varphi
- \frac{1}{2} \tan\frac{\theta}{2} \sigma_3 \right), \nonumber \\
&&J_3 = - i \p_\varphi - \frac{\sigma_3}{2},
\end{eqnarray}
and is given by
\begin{equation}
H(0) = \frac{1}{2 \Theta}\left(\vec J^{\, 2} - \frac{1}{4}\right) +  E(\vec p),
\end{equation}
such that the energy spectrum takes the form
\begin{equation}
E_j(0) = \frac{1}{2 \Theta} \left[j(j + 1) - \frac{1}{4}\right] + E(\vec p).
\end{equation}
Here $j$ is a half-integer. In this case, each state is $2(2j+1)$-fold 
degenerate because the fermion sectors $+$ and $-$ cost the same energy. The 
corresponding wave functions with half-integer angular momentum are monopole 
harmonics \cite{Fie44,Wu76}. In particular, the ground state wave functions are
\begin{eqnarray}
&&Y^\pm_{\frac{1}{2},\pm \frac{1}{2}}(\theta,\varphi) = 
\frac{1}{\sqrt{2 \pi}} \sin\frac{\theta}{2} \exp(\pm i \varphi), \nonumber \\
&&Y^\pm_{\frac{1}{2},\mp \frac{1}{2}}(\theta,\varphi) = 
\frac{1}{\sqrt{2 \pi}} \cos\frac{\theta}{2}.
\end{eqnarray}
It should be noted that $Y^+_{\frac{1}{2},\frac{1}{2}}(\theta,\varphi)$ and 
$Y^-_{\frac{1}{2},-\frac{1}{2}}(\theta,\varphi)$ have coordinate singularities 
at $\theta = \pi$ related to the Dirac string. Following Wu and Yang 
\cite{Wu76}, one can avoid the coordinate singularity by introducing different 
coordinate patches glued together by gauge transformations.

The Hamiltonian with $\lambda \neq 0$ takes the form
\begin{equation}
H(\lambda) = H(0) + \frac{1}{2 \Theta}
\left(\lambda C + \frac{1}{2} \lambda^2 \right),
\end{equation}
and still commutes with $\vec J$ of eq.(\ref{J}). Here
\begin{eqnarray}
C&=&- i \left(\sin\varphi \p_\theta + 
\frac{\cos\varphi}{\sin\theta} \p_\varphi -
\frac{1}{2} \sin\varphi \tan\frac{\theta}{2}\right) \sigma_1 \nonumber \\
&+&i \left(\cos\varphi \p_\theta - \frac{\sin\varphi}{\sin\theta} \p_\varphi -
\frac{1}{2} \cos\varphi \tan\frac{\theta}{2}\right) \sigma_2,
\end{eqnarray}
and $[C,\vec J] = 0$. Using $C^2 = \vec J^{\, 2} + \frac{1}{4}$ one obtains the
energy spectrum
\begin{equation}
E_j(\lambda) = \frac{1}{2 \Theta} \left[j'(j' + 1)
+ \frac{\lambda^2 - 1}{4}\right] + E(\vec p),
\end{equation}
with $j' = j \pm \frac{\lambda}{2}$ and $j$ again being a half-integer. For 
$\lambda \neq 0$ the fermion sectors $+$ and $-$ get mixed and the previously 
degenerate $2(2j+1)$ states are now split into two groups of $2j+1$ degenerate 
states. Interestingly, for $\lambda = \pm 1$ the monopole field strength of 
eq.(\ref{F}) vanishes and $E_j(\pm 1) = \frac{1}{2 \Theta} j'(j' + 1)$ with 
$j' = j \pm \frac{1}{2}$. In that case, the rotor spectrum looks like the one 
of eq.(\ref{Energy}) although the angular momentum $j$ is now a half-integer.

Let us now consider QCD with two massless flavors and thus with a spontaneously
broken $SU(2)_L \times SU(2)_R$ chiral symmetry. When the theory is put in a
finite spatial volume $V$, as it is the case in numerical simulations of
lattice QCD, the chiral order parameter $U(t) \in SU(2)$ describes a quantum 
rotor with the Lagrangian
\begin{equation}
{\cal L} = \frac{\Theta}{4} \mbox{Tr}\left[\p_t U^\dagger \p_t U\right].
\end{equation}
At tree-level the moment of inertia is given by $\Theta = F_\pi^2 V$ where
$F_\pi$ is the pion decay constant. The corresponding Hamiltonian is the 
Laplacian on the sphere $S^3$. The QCD rotor spectrum has been derived by 
Leutwyler \cite{Leu87} in the $\delta$-expansion of chiral perturbation theory 
\cite{Gas85} as
\begin{equation}
\label{EnergyQCD}
E_l = \frac{j_L(j_L + 1) + j_R(j_R + 1)}{\Theta} =
\frac{l(l + 2)}{2 \Theta}.
\end{equation}
In this case, $j_L = j_R$ with $l = j_L + j_R \in \{0,1,2,...\}$ and each state
is $(2 j_L + 1)(2 j_R + 1) = (l+1)^2$-fold degenerate. The low-energy dynamics 
of nucleons and pions is described by baryon chiral perturbation theory
\cite{Gas88,Jen91,Jen92,Ber92}. When a nucleon with small momentum 
$\vec p = |\vec p| \vec e_p$ is propagating in the finite volume the Lagrangian
reads
\begin{eqnarray}
{\cal L}&=&\frac{\Theta}{4} \mbox{Tr}\left[\p_t U^\dagger \p_t U\right] 
\nonumber \\
&+&\Psi^\dagger \left[E(\vec p) - i \p_t - i v_t - 
i \lambda (\vec \sigma \cdot \vec e_p) a_t\right] \Psi. 
\end{eqnarray}
Here $\Psi(t)$ is a Pauli spinor with a flavor index distinguishing
protons and neutrons and $\frac{\vec \sigma}{2}$ is the nucleon spin. At tree
level $E(\vec p) = M + \vec p^{\, 2}/2 M$ and $\lambda = g_A |\vec p|/M$, 
where $M$ is the mass and $g_A$ is the axial vector coupling of the nucleon. 
As for the antiferromagnet, the parameters $\Theta$, $E(\vec p)$, and $\lambda$
get renormalized by the coupling to non-zero momentum pion modes. In the QCD 
case $u^2 = U$ and
\begin{equation}
v_t = \frac{1}{2}\left(u \p_t u^\dagger + u^\dagger \p_t u\right), \ 
a_t = \frac{1}{2i}\left(u \p_t u^\dagger - u^\dagger \p_t u\right).
\end{equation}
Parameterizing
\begin{eqnarray}
&&U(t) = \cos\alpha(t) + i \sin\alpha(t) \vec e_\alpha(t) \cdot \vec\tau,
\nonumber \\
&&\vec e_\alpha(t) = (\sin\theta(t) \cos\varphi(t),
\sin\theta(t) \sin\varphi(t),\cos\theta(t)), \nonumber \\
&&\vec e_\theta(t) = (\cos\theta(t) \cos\varphi(t),
\cos\theta(t) \sin\varphi(t),- \sin\theta(t)), \nonumber \\
&&\vec e_\varphi(t) = (- \sin\varphi(t),\cos\varphi(t),0),
\end{eqnarray}
one obtains
\begin{eqnarray}
v_t&=&i \sin^2\frac{\alpha}{2}\left(\p_t \theta \ \vec e_\varphi -
\sin\theta \ \p_t \varphi \ \vec e_\theta\right) \cdot \vec \tau, \nonumber \\
a_t&=&\left(\frac{\p_t \alpha}{2} \vec e_\alpha + 
\sin\alpha \frac{\p_t \theta}{2} \vec e_\theta +
\sin\alpha \sin\theta \frac{\p_t \varphi}{2} \vec e_\varphi\right) 
\cdot \vec \tau. \nonumber \\
\end{eqnarray}
Here $\vec \tau$ are the Pauli matrices for isospin.

The resulting Hamilton operator takes the form
\begin{eqnarray}
&&H(\lambda) = - \frac{1}{2 \Theta} \left\{\frac{1}{\sin^2\alpha}
(\p_\alpha - A_\alpha)[\ \sin^2\alpha (\p_\alpha - A_\alpha)]\right. 
\nonumber \\
&&+ \frac{1}{\sin^2\alpha \sin\theta}
(\p_\theta - A_\theta)[\ \sin\theta (\p_\theta - A_\theta)] \nonumber \\
&&+ \left.\frac{1}{\sin^2\alpha \sin^2\theta} (\p_\varphi - A_\varphi)^2
\right\} + E(\vec p),
\end{eqnarray}
with the non-Abelian vector potential
\begin{eqnarray}
A_\alpha&=&i \frac{\lambda}{2} (\vec \sigma \cdot \vec e_p) 
\vec e_\alpha \cdot \vec \tau, 
\nonumber \\
A_\theta&=&i \left(\sin^2\frac{\alpha}{2} \ \vec e_\varphi + 
\frac{\lambda}{2} (\vec \sigma \cdot \vec e_p) 
\sin\alpha \ \vec e_\theta\right) \cdot \vec \tau, 
\nonumber \\
A_\varphi&=&i \left(- \sin^2\frac{\alpha}{2} \sin\theta \ \vec e_\theta \right.
\nonumber \\
&+&\left.\frac{\lambda}{2} (\vec \sigma \cdot \vec e_p) 
\sin\alpha \sin\theta \ \vec e_\varphi\right) \cdot \vec \tau,
\end{eqnarray}
and the corresponding field strength
\begin{eqnarray}
\label{FQCD}
F_{\alpha\theta}&=&i \frac{1 - \lambda^2}{2} \sin\alpha \ \vec e_\varphi 
\cdot \vec \tau, \nonumber \\
F_{\theta\varphi}&=&i \frac{1 - \lambda^2}{2} \sin^2\alpha \ \sin\theta \ 
\vec e_\alpha \cdot \vec \tau, \nonumber \\
F_{\varphi\alpha}&=&i \frac{1 - \lambda^2}{2} \sin\alpha \ \sin\theta \
\vec e_\theta \cdot \vec \tau.
\end{eqnarray}
This Berry gauge field is a non-Abelian analog of the Abelian monopole field 
we encountered for the antiferromagnet. The non-Abelian gauge field again has
a coordinate singularity, in this case at $\alpha = \pi$, which corresponds to 
a Dirac string going through the south pole of $S^3$. This indicates that there
is a non-Abelian magnetic monopole at the center of $S^3$.

The generators of $SU(2)_L \otimes SU(2)_R$ take the form
\begin{eqnarray}
\vec J_L&=&\frac{1}{2}(\vec J - \vec K), \
\vec J_R = \frac{1}{2}(\vec J + \vec K), \nonumber \\
J_\pm&=&\exp(\pm i \varphi)\left(\pm \ \p_\theta + 
i \cot\theta \ \p_\varphi\right) + \frac{\tau_\pm}{2}, \nonumber \\
J_3&=&- i \p_\varphi + \frac{\tau_3}{2}, \nonumber \\ 
K_\pm&=&\exp(\pm i \varphi)\left(i \sin\theta \ \p_\alpha + 
i \cot\alpha \cos\theta \ \p_\theta \mp 
\frac{\cot\alpha}{\sin\theta} \p_\varphi\right. \nonumber \\
&\mp&\left.\frac{i}{2} \tan\frac{\alpha}{2} \ \vec e_\theta \cdot \vec \tau +
\frac{1}{2} \tan\frac{\alpha}{2} \cos\theta \ \vec e_\varphi \cdot \vec \tau
\right), \nonumber \\
K_3&=&i \left(\cos\theta \ \p_\alpha - \cot\alpha \sin\theta \ \p_\theta\right)
\nonumber \\
&-&\frac{1}{2} \tan\frac{\alpha}{2} \sin\theta \ 
\vec e_\varphi \cdot \vec \tau.
\end{eqnarray}
The Hamiltonian $H(0)$ (with $\lambda = 0$) can be written as
\begin{equation}
H(0) = \frac{1}{2 \Theta}\left(\vec J^{\, 2} + \vec K^{\, 2} - 
\frac{3}{4}\right) + E(\vec p),
\end{equation}
such that the energy spectrum takes the form
\begin{equation}
E_j(0) = \frac{1}{2 \Theta} \left[j (j + 2) - \frac{1}{2}\right]
+ E(\vec p).
\end{equation}
In this case, $j_L = j_R \pm \frac{1}{2}$ and $j = j_L + j_R \in 
\{\frac{1}{2},\frac{3}{2},...\}$. Each state is 
$2 (j + \frac{1}{2})(j + \frac{3}{2})$-fold degenerate because the states with
spin up and spin down cost the same energy.

The Hamiltonian with $\lambda \neq 0$ can be written as
\begin{equation}
H(\lambda) = H(0) + \frac{1}{2 \Theta} \left(\lambda C + 
\frac{3}{4} \lambda^2\right),
\end{equation}
and it still commutes with $\vec J$ and $\vec K$. Here
\begin{eqnarray}
C&=&i (\vec \sigma \cdot \vec e_ p) \left(\vec e_\alpha \p_\alpha + 
\frac{1}{\sin\theta} \vec e_\theta \p_\theta \right. \nonumber \\
&+&\left.\frac{1}{\sin\alpha \sin\theta} \vec e_\varphi \p_\varphi -
\tan\frac{\alpha}{2} \vec e_\alpha\right) \cdot \vec \tau,
\end{eqnarray}
and $[C,\vec J] = [C,\vec K] = 0$. Using 
$C^2 = \vec J^{\, 2} + \vec K^{\, 2} + \frac{3}{4}$ one finally obtains the 
energy spectrum
\begin{equation}
E_j(\lambda) = \frac{1}{2 \Theta} \left[j' (j' + 2) + 
\frac{\lambda^2 - 1}{2}\right] + E(\vec p),
\end{equation}
with $j' = j \pm \frac{\lambda}{2}$, where $\pm$ refers to the spin eigenstates
of $\vec\sigma \cdot \vec e_p$ with eigenvalues $\pm 1$. Thus we see that for 
$\lambda \neq 0$ the degeneracy is partly lifted and there are now two groups 
of $(j + \frac{1}{2})(j + \frac{3}{2})$-fold degenerate states. Remarkably, for
$\lambda = \pm 1$ the non-Abelian field strength of eq.(\ref{FQCD}) vanishes 
and $E_j(\pm 1) = \frac{1}{2 \Theta} j' (j' + 2)$ with 
$j' = j \pm \frac{1}{2}$. Just as for an antiferromagnet with 
$\lambda = \pm 1$, the QCD rotor spectrum then looks like the one of
eq.(\ref{EnergyQCD}) although the system now has fermion number one.

The present study in the $\delta$-regime complements other investigations of 
finite volume effects in the one-nucleon sector of QCD in the $p$- 
\cite{Ali04,Beane04,Col06}, $\epsilon$- \cite{Bed05}, and $\epsilon'$-regimes 
\cite{Det04} of chiral perturbation theory. A comparison of numerical lattice 
QCD data with finite volume predictions of chiral perturbation theory may lead 
to an accurate determination of low-energy parameters including $F_\pi$ and 
some of the Gasser-Leutwyler coefficients. Before one can do this in the 
$\delta$-regime, one must match the volume-dependent parameters $\Theta$, 
$E(\vec p)$, and $\lambda$ of the effective quantum mechanics to those of the 
infinite volume effective theory. 

While it is difficult to simulate QCD in the chiral limit, simulations of a
single hole in the $t$-$J$ model with an exact $SU(2)_s$ spin symmetry are 
possible using efficient cluster algorithms. Again, before one can extract the 
parameters of the systematic effective theory of magnons and holes \cite{Bru06}
from a comparison with numerical data, it will be necessary to match the
volume-dependent parameters $\lambda$ and $E(\vec p)$ of the effective quantum
mechanics to those of the infinite volume effective theory, e.g.\ at one loop.
However, even without performing this matching calculation, one can check if
indeed $\lambda = 0$, as predicted for the $t$-$J$ model in \cite{Bru06}.

The effects discussed here are not limited to antiferromagnets or QCD, but
arise for any finite system with a spontaneously broken continuous global 
symmetry (unless the order parameter is conserved). When 
a symmetry group $G$ breaks spontaneously to a subgroup $H$, the corresponding 
order parameter takes values in the coset space $G/H$. In a finite volume the 
symmetry is restored by a slow rotation of the order parameter. When a single 
fermion is added to the system, one expects Berry phases resulting from 
monopole gauge fields residing on the manifold $G/H$, with characteristic 
effects on the rotor spectrum. It would be interesting to work 
out these effects for general $G$ and $H$ and for an arbitrary fermion 
representation. A non-trivial case of physical interest is QCD with three 
massless flavors for which $G/H = SU(3)$ and the baryons transform as flavor 
octets. 

We have benefitted from discussions with G.\ Colangelo, S.\ D\"urr, 
C.\ Haefeli, P.\ Hasenfratz, and F.\ Niedermayer. This work was supported in 
part by the U.S.\ Department of Energy grant DE-FG02-05ER41368 and by the
Schweizerischer Nationalfonds.

\end{document}